\documentclass[preprint,12pt,authoryear]{elsarticle}

\usepackage{amssymb}
\usepackage{amsmath}
\usepackage{placeins}
\usepackage{booktabs}
\usepackage{graphicx}
\usepackage{adjustbox}

\usepackage{url}
\usepackage{lineno}

\journal{Medical Image Analysis}

\begin{document}

\begin{frontmatter}

\title{STRIDE: Spatial-Temporal Representation for Interval-conditioned Disease Evolution in Longitudinal Glioblastoma MRI}

\author[aff1]{Wenhao Guo\fnref{equal}}
\author[aff2]{Changchang Yin\fnref{equal}}
\author[aff3]{Pierre Giglio}
\author[aff2]{Weidan Cao}
\author[aff1,aff2]{Ping Zhang\corref{cor1}}
\ead{zhang.10631@osu.edu}
\author[aff1]{Golrokh Mirzaei\corref{cor2}}
\ead{mirzaei.4@osu.edu}

\fntext[equal]{These authors contributed equally to this work.}
\cortext[cor1]{Corresponding author.}
\cortext[cor2]{Corresponding author.}

\affiliation[aff1]{
    organization={Department of Computer Science and Engineering, The Ohio State University},
    city={Columbus},
    state={OH},
    country={USA}}

\affiliation[aff2]{
    organization={Department of Biomedical Informatics, The Ohio State University},
    city={Columbus},
    state={OH},
    country={USA}}

\affiliation[aff3]{
    organization={Department of Neurology, The Ohio State University},
    city={Columbus},
    state={OH},
    country={USA}}

\begin{abstract}
Glioblastoma (GBM), an aggressive primary brain tumor, is routinely
monitored with longitudinal MRI after treatment. Distinguishing stable disease
(SD), pseudoprogression (PsP), and true progression (TP) remains challenging
because these states can show overlapping MRI appearances despite different
temporal trajectories. Existing longitudinal methods still face challenges
in modeling scan-specific spatial variability, variable follow-up intervals,
and complementary information from the observed follow-up state and its
longitudinal change. We propose \textbf{STRIDE}, a framework for spatial-temporal
representation of interval-conditioned disease evolution that takes paired
post-treatment MRI scans and their inter-scan interval as input and predicts
SD, PsP, or TP. The lesion-prior-guided spatial representation combines
SoftGate and an Adaptive-window Hierarchical Transformer (AWHT) to emphasize
lesion-related regions while preserving surrounding context. The
time-conditioned latent transition uses pair-level context and the actual
inter-scan interval to estimate interval-dependent representation changes
between visits. The observed--transition fusion integrates the
transition-estimated follow-up representation with the directly observed
follow-up representation to jointly characterize the follow-up state and its
longitudinal change. BraTS2024 is used to develop and evaluate the
lesion-prior generator, while longitudinal pretraining on LUMIERE supports
transfer before downstream adaptation to Burdenko. On the Burdenko three-class
task, STRIDE achieves a macro ROC--AUC of 0.816 and a macro F1-score of 0.796.
These results support its potential for more reliable longitudinal
post-treatment GBM state assessment.
\end{abstract}



\end{frontmatter}


\section{Introduction}
\label{sec:intro}

Glioblastoma (GBM) is the most aggressive primary brain tumor in adults and
remains a major challenge in neuro-oncology. Despite advances in surgery,
radiotherapy, and chemotherapy, tumor recurrence and progression after
treatment remain common, leading to poor clinical outcomes
~\citep{stupp2005radiotherapy}. Longitudinal magnetic resonance imaging (MRI)
is routinely used to monitor treatment response and guide clinical decisions,
with standardized criteria such as RANO and its updated versions supporting
imaging-based response assessment
~\citep{wen2010updated,okada2015immunotherapy,okada2018immunotherapy,wen2023rano}.
However, reliable differentiation among stable disease (SD),
pseudoprogression (PsP), and true progression (TP) remains challenging because
post-treatment imaging appearances can overlap while the corresponding disease
trajectories differ
~\citep{wen2010updated,taal2008incidence,brandes2008mgmt}. Distinguishing PsP
from TP is especially difficult during early post-treatment follow-up, making
longitudinal assessment important for reliable clinical interpretation
~\citep{taal2008incidence,brandes2008mgmt,young2023pseudoprogression}.

Recent artificial intelligence approaches have demonstrated potential for
automated GBM progression assessment, including radiomics and classical machine
learning
~\citep{turk2022comparative,jang2018prediction,kim2019incorporating,
elshafeey2019multicenter,akbari2020histopathology}, volumetric and multimodal
deep learning
~\citep{moassefi2022deep,li2025deep}, Transformer-based architectures
~\citep{guo2025timepoint}, and self-supervised representation learning
~\citep{gomaa2025self}. Despite this progress, systematic reviews and
meta-analyses have reported substantial variability in diagnostic performance
across cohorts
~\citep{palavani2025artificial,kotochinsky2026machine}. One persistent
challenge is that a single MRI examination provides only a snapshot of the
disease state and does not directly capture how the disease changes between
visits. This has motivated recurrent models
~\citep{siakallis2021longitudinal,cui2019rnn}, Siamese and difference-based
learning
~\citep{li2020siamese,zheng2025deep}, and temporal attention and
Transformer-based longitudinal approaches
~\citep{aghajanian2025longitudinal,wang2024ordinal,
nguyen2023clinically,sun2024lomia,zhang2025glomia,zheng2026whole,
tarek2026trace}. However, incorporating multiple visits does not by itself
resolve three key modeling challenges.

First, progression-related abnormalities vary in location and spatial extent
across MRI scans, while existing longitudinal representations remain
limited in adapting to such scan-specific spatial variability
~\citep{guo2025timepoint,gomaa2025self}. Second, follow-up intervals vary
substantially across patients, yet the actual inter-scan interval remains
insufficiently incorporated in existing longitudinal modeling of post-treatment
GBM
~\citep{chen2018neural,rubanova2019latent,che2018recurrent,
ghazi2019training,mouadden2025conditional,de2019gru,liu2025imageflownet,
tarek2026trace}. Third, the observed follow-up state and the longitudinal
change leading to that state provide complementary information, while existing
longitudinal models remain limited in jointly characterizing these two aspects
of disease evolution
~\citep{cui2019rnn,li2020siamese,sun2024lomia,zhang2025glomia,
zheng2026whole,tarek2026trace}.

To address these challenges, we propose \textbf{STRIDE}, a spatial-temporal
representation framework for interval-conditioned disease evolution that takes
paired post-treatment MRI scans and their inter-scan interval as input and
predicts SD, PsP, or TP. For each MRI scan, the lesion-prior-guided spatial
representation uses segmentation-derived structural guidance
~\citep{ronneberger2015u,isensee2021nnu,hatamizadeh2022unetr,
Hatamizadeh2022SwinUNETR,mehta2018net,fang2019attention,gao2026learning}
through a mask-conditioned SoftGate to emphasize lesion-supported features
while retaining surrounding context. An Adaptive-window Hierarchical
Transformer (AWHT) then combines scan-adaptive multi-scale localization with
hierarchical encoding to produce a visit-level spatial representation. The
spatial representations from the two visits are subsequently used to construct
pair-level longitudinal context. The time-conditioned latent transition
combines the earlier-visit representation, pair-level context, and the actual
inter-scan interval to estimate interval-dependent representation changes and
produce a transition-estimated follow-up representation. The
observed--transition fusion then integrates the earlier-visit representation,
the directly observed and transition-estimated follow-up representations, and
their associated longitudinal information to jointly characterize the
follow-up state and its longitudinal change. Because annotated longitudinal
GBM cohorts remain limited, longitudinal pretraining is additionally employed
to improve representation transfer before downstream adaptation
~\citep{zhou2021models,tang2022self,huo2025time,huang2026longitudinal,
gomaa2025self,suter2022lumiere}.

Our contributions are summarized as follows:

\begin{enumerate}

\item We propose STRIDE, a longitudinal Transformer framework for three-class
post-treatment GBM clinical-state classification that jointly addresses
scan-specific spatial heterogeneity, variable inter-scan intervals, and
complementary state--change information.

\item We introduce lesion-prior-guided spatial representation that combines
structural guidance with scan-adaptive multi-scale localization to emphasize
lesion-related regions while preserving surrounding context.

\item We develop a time-conditioned latent transition that incorporates
pair-level context and the actual inter-scan interval to explicitly model
interval-dependent representation changes between visits.

\item We design an observed--transition fusion that integrates directly
observed and transition-estimated follow-up representations to capture
complementary state--change information for SD, PsP, and TP classification.

\end{enumerate}

\section{Proposed Method}
\label{sec:methods}


\subsection{Overview of STRIDE Framework}
\label{sec:method_overview}

\begin{figure*}[t]
    \centering
    \includegraphics[width=1.1\linewidth]{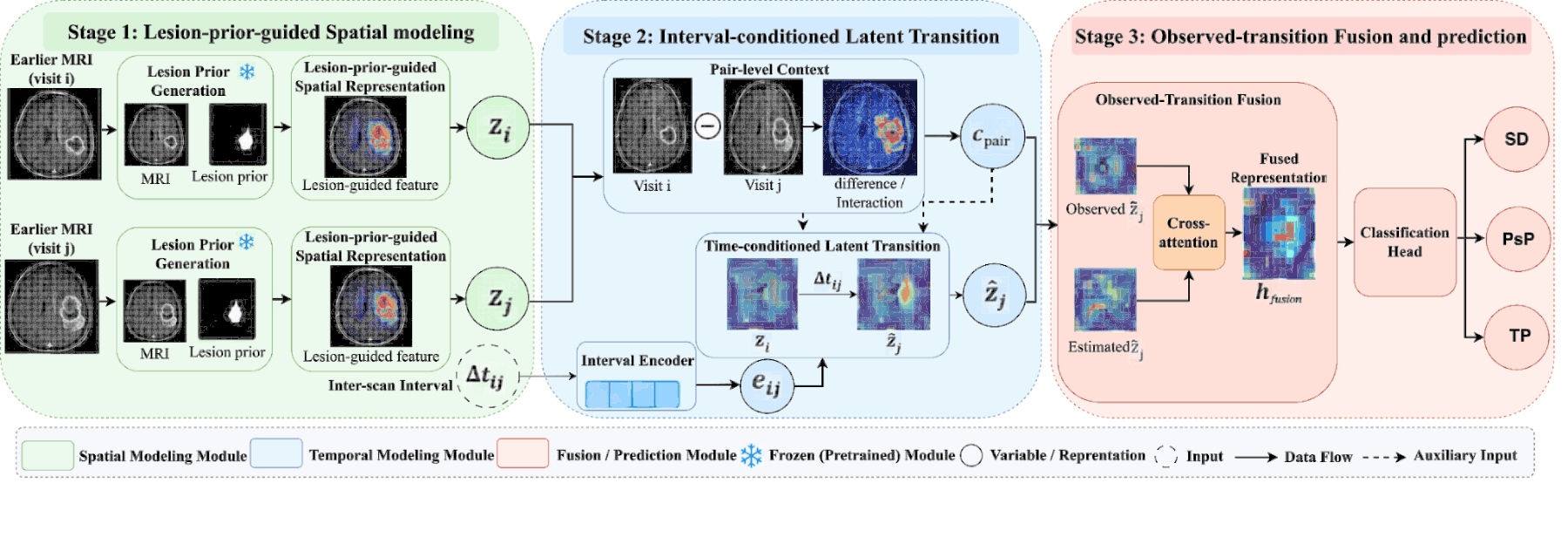}
    \caption{
    \textbf{Overall architecture of STRIDE.}
Given paired longitudinal MRI scans and their inter-scan interval,
segmentation-derived lesion priors first provide structural guidance for
lesion-prior-guided spatial representation. The resulting visit-specific
representations are used to construct pair-level context, while the inter-scan
interval is encoded as temporal information. These representations are then
processed by the time-conditioned latent transition to characterize
interval-dependent latent changes and obtain a transition-estimated follow-up
representation. Finally, the observed and transition-estimated longitudinal
representations are jointly fused for three-class classification of stable
disease (SD), pseudoprogression (PsP), and true progression (TP).
    }
    \label{fig:framework}
\end{figure*}

STRIDE consists of three major components:
lesion-prior-guided spatial representation,
time-conditioned latent transition, and
observed--transition fusion. These components
progressively extract visit-specific spatial representations, characterize
interval-dependent changes between scans, and integrate observed and
transition-estimated longitudinal information for three-class clinical-state
classification. The overall architecture is illustrated in
Figure~\ref{fig:framework}.

\subsection{Problem Formulation}
\label{sec:problem_formulation}

For a patient with $N$ longitudinal MRI scans, we denote the imaging sequence
as $\mathcal{X}=\{(X_i,t_i)\}_{i=1}^{N}$, where
$X_i\in\mathbb{R}^{M\times H\times W\times D}$ is the multimodal MRI scan
acquired at visit $i$, $t_i$ is its acquisition time, $M$ denotes the number
of imaging modalities, and $H$, $W$, and $D$ denote the spatial dimensions.

Although a patient may have multiple MRI scans, STRIDE operates on visit
pairs. For visits $i<j$, the corresponding inter-scan interval is
$\Delta t_{ij}=t_j-t_i$, and each sample is represented by
$(X_i,X_j,\Delta t_{ij})$. For downstream classification, each pair is
assigned the patient's curated clinical-state label
$c_{ij}\in\{\mathrm{SD},\mathrm{PsP},\mathrm{TP}\}$ and its corresponding
one-hot vector $\mathbf{y}_{ij}\in\{0,1\}^{3}$. Given a longitudinal pair,
STRIDE predicts

\begin{equation}
\mathbf{p}_{ij}
=
F_{\theta}
\left(
X_i,X_j,\Delta t_{ij}
\right),
\qquad
\sum_{k=1}^{3}p_{ij,k}=1,
\label{eq:task_definition}
\end{equation}
where $\mathbf{p}_{ij}\in[0,1]^3$, and its three elements correspond to
SD, PsP, and TP, respectively. The predicted clinical state is
$\hat{c}_{ij}=\arg\max_k p_{ij,k}$.


\subsection{Lesion-prior-guided Spatial Representation}
\label{sec:spatial_representation}

Post-treatment abnormalities can be concentrated within localized lesion
regions, while surrounding tissue can still contain information relevant to
distinguishing SD, PsP, and TP. Treating all spatial locations equally makes
lesion-related signals less prominent, whereas restricting representation
learning to a hard lesion mask removes useful contextual features. We therefore
combine segmentation-derived structural guidance with soft lesion-aware
modulation and adaptive spatial representation learning.

\begin{figure*}[t]
    \centering
    \includegraphics[width=1\linewidth]{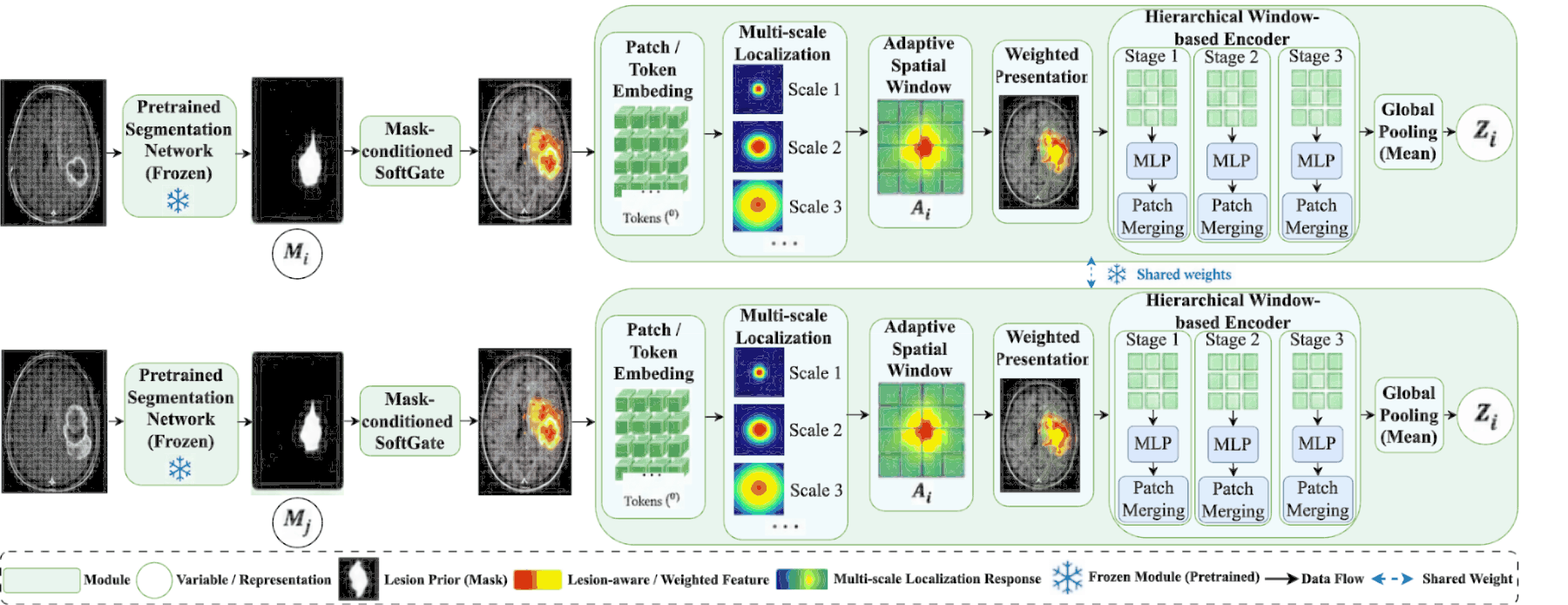}
    \caption{
\textbf{Lesion-prior-guided spatial representation.}
For each visit, a segmentation-derived lesion prior provides structural
guidance to the mask-conditioned SoftGate, which emphasizes lesion-supported
features while retaining surrounding contextual information. The resulting
representation is processed by the Adaptive-window Hierarchical Transformer
(AWHT). AWHT constructs an input-dependent continuous spatial window from
multi-scale localization responses and subsequently encodes the weighted
representation through hierarchical window-based self-attention, producing
the visit-level representation $\mathbf{z}_v$. The spatial representation
pathway is shared across longitudinal visits.
}
    \label{fig:spatial_module}
\end{figure*}

For each visit, a frozen segmentation network provides a lesion prior.
SoftGate uses this prior to modulate image features, and the resulting
representation is processed by the Adaptive-window Hierarchical Transformer
(AWHT) with scan-adaptive spatial emphasis. The module is illustrated in
Figure~\ref{fig:spatial_module}.

\subsubsection{Segmentation-derived Lesion Prior}

For visit $v$, a pretrained segmentation network generates a voxel-wise
abnormal-region probability map
$M_v=S(X_v)$, where
$M_v\in[0,1]^{1\times H\times W\times D}$. The segmentation network remains
frozen during longitudinal learning, and its output is used only as auxiliary
structural guidance for subsequent spatial representation learning.


\subsubsection{Mask-conditioned SoftGate}

SoftGate applies the lesion prior as a soft emphasis signal while preserving
the original feature representation.

Let
$\mathbf{F}_v\in
\mathbb{R}^{H_F\times W_F\times D_F\times C_F}$
denote the visit-specific feature map used by SoftGate. After spatial alignment,
the lesion prior is projected by $\phi(\cdot)$ to a representation compatible
with $\mathbf{F}_v$. A continuous gating map is then obtained as

\begin{equation}
\mathbf{G}_v
=
\sigma
\left(
\mathbf{F}_v
\odot
\phi(M_v)
\right),
\label{eq:softgate_map}
\end{equation}
where $\phi(\cdot)$ denotes a learnable prior projection,
$\sigma(\cdot)$ denotes the sigmoid activation, and
$\odot$ denotes element-wise multiplication.

The gate is applied through residual modulation:

\begin{equation}
\widetilde{\mathbf{F}}_v
=
\mathbf{F}_v
+
\alpha\,
\mathbf{F}_v\odot\mathbf{G}_v,
\label{eq:softgate}
\end{equation}
where $\alpha\in[0,1]$ controls the strength of lesion-aware modulation.
In all experiments, $\alpha$ was fixed to 0.5.


\subsubsection{Adaptive-window Hierarchical Transformer}
\label{sec:awht}

To accommodate variations in lesion location and spatial extent, AWHT
constructs an input-dependent continuous spatial window from multi-scale
image features before hierarchical window-based Transformer encoding. It
contains an input-adaptation stem, multi-scale adaptive spatial-window
estimation, and hierarchical 3D window-based representation encoding. Here,
\emph{adaptive window} refers to the examination-specific continuous
spatial weighting field, while the local self-attention window remains fixed
at $8\times8\times8$ tokens.

\paragraph{Input adaptation}

The SoftGate-modulated representation is first projected into a common feature
space using two consecutive $3\times3\times3$ convolutional blocks:

\begin{equation}
\begin{aligned}
\mathbf{U}_v
&=
\operatorname{Dropout3D}
\left[
\psi_2
\left(
\psi_1(\widetilde{\mathbf{F}}_v)
\right)
\right],
\\
\psi_r(\mathbf{F})
&=
\operatorname{GELU}
\left[
\operatorname{BN}_r
\left(
\operatorname{Conv}^{3\times3\times3}_r(\mathbf{F})
\right)
\right].
\end{aligned}
\label{eq:awht_input}
\end{equation}
where
$\mathbf{U}_v\in
\mathbb{R}^{H_F\times W_F\times D_F\times C_0}$
and $C_0$ is the stem feature dimension.

\paragraph{Multi-scale adaptive spatial window}

Because abnormalities can occupy substantially different spatial extents,
localization cues obtained from a single receptive-field scale are
insufficient. We therefore use three parallel localization branches operating
at scales $\mathcal{S}=\{1,2,4\}$:

\begin{equation}
\mathbf{L}^{(s)}_v
=
\operatorname{Conv}^{1\times1\times1}_{s,2}
\left[
\operatorname{ReLU}
\left(
\operatorname{BN}_{s}
\left[
\operatorname{Conv}^{3\times3\times3}_{s,1}
(\mathbf{U}_v)
\right]
\right)
\right],
\label{eq:awht_multiscale}
\end{equation}
where the first convolution uses stride $s\in\mathcal{S}$. Larger strides
provide localization cues over broader effective receptive fields.

The scale-specific responses are subsequently aligned to the spatial
resolution of $\mathbf{U}_v$ and aggregated to obtain the adaptive spatial
window

\begin{equation}
\mathbf{A}_v
=
\sigma
\left[
\frac{1}{|\mathcal{S}|}
\sum_{s\in\mathcal{S}}
\mathcal{I}_{s}
\left(
\mathbf{L}^{(s)}_v
\right)
\right],
\label{eq:awht_window}
\end{equation}
where
$\mathbf{A}_v\in
[0,1]^{H_F\times W_F\times D_F\times1}$ and
$\mathcal{I}_{s}(\cdot)$ denotes trilinear interpolation to the common spatial
resolution.

The adaptive spatial window is then applied through element-wise weighting,
$\widetilde{\mathbf{U}}_v=\mathbf{U}_v\odot\mathbf{A}_v$, where
$\mathbf{A}_v$ is broadcast across feature channels and
$\widetilde{\mathbf{U}}_v\in
\mathbb{R}^{H_F\times W_F\times D_F\times C_0}$.
This produces an examination-specific weighted feature volume for hierarchical
representation encoding.

\paragraph{Hierarchical window-based representation encoding}

The weighted feature volume $\widetilde{\mathbf{U}}_v$ is divided into
non-overlapping $4\times4\times4$ patches and projected to an initial embedding
dimension $C_0=96$:

\begin{equation}
\mathbf{Z}^{(0)}_v
=
\operatorname{PatchEmbed}
\left(
\widetilde{\mathbf{U}}_v
\right),
\label{eq:awht_patch}
\end{equation}
where
$\mathbf{Z}^{(0)}_v\in
\mathbb{R}^{N_0\times C_0}$ and
$N_0$ denotes the number of initial 3D patch tokens.

The hierarchical encoder contains four stages with block depths
$(2,2,2,2)$ and attention-head configurations $(3,6,12,24)$. Let
$\mathbf{Z}^{(l,b-1)}_v\in\mathbb{R}^{N_l\times C_l}$ denote the input
representation to block $b$ of stage $l$, where $N_l$ is the number of tokens
and $C_l$ is the corresponding embedding dimension. Within each stage,
regular window-based self-attention and shifted-window self-attention are
alternated:

\begin{align}
\widehat{\mathbf{Z}}^{(l,b)}_v
&=
\mathbf{Z}^{(l,b-1)}_v
+
\mathcal{A}^{(l,b)}
\left[
\operatorname{LN}
\left(
\mathbf{Z}^{(l,b-1)}_v
\right)
\right],
\\
\mathbf{Z}^{(l,b)}_v
&=
\widehat{\mathbf{Z}}^{(l,b)}_v
+
\operatorname{MLP}
\left[
\operatorname{LN}
\left(
\widehat{\mathbf{Z}}^{(l,b)}_v
\right)
\right],
\label{eq:awht_block}
\end{align}
where
$\widehat{\mathbf{Z}}^{(l,b)}_v,
\mathbf{Z}^{(l,b)}_v
\in\mathbb{R}^{N_l\times C_l}$,
and $\mathcal{A}^{(l,b)}$ alternates between regular and shifted-window
self-attention. Patch merging between consecutive stages reduces the number of
spatial tokens while increasing the embedding dimension, producing
$\mathbf{Z}^{(l+1,0)}_v
\in\mathbb{R}^{N_{l+1}\times C_{l+1}}$.

After the final hierarchical stage, global average pooling and normalization
produce the visit-level representation

\begin{equation}
\mathbf{z}_v
=
\operatorname{LN}
\left[
\operatorname{GAP}
\left(
\mathbf{Z}^{(L-1)}_v
\right)
\right].
\label{eq:awht_output}
\end{equation}
where $\mathbf{z}_v\in\mathbb{R}^{Z}$.

Applying the same spatial encoder to both examinations yields
$\mathbf{z}_{i}$ and $\mathbf{z}_{j}\in\mathbb{R}^{Z}$ for subsequent
longitudinal modeling.


\subsection{Time-conditioned Latent Transition}
\label{sec:temporal_transition}

\begin{figure*}[t]
    \centering
    \includegraphics[width=0.98\linewidth]{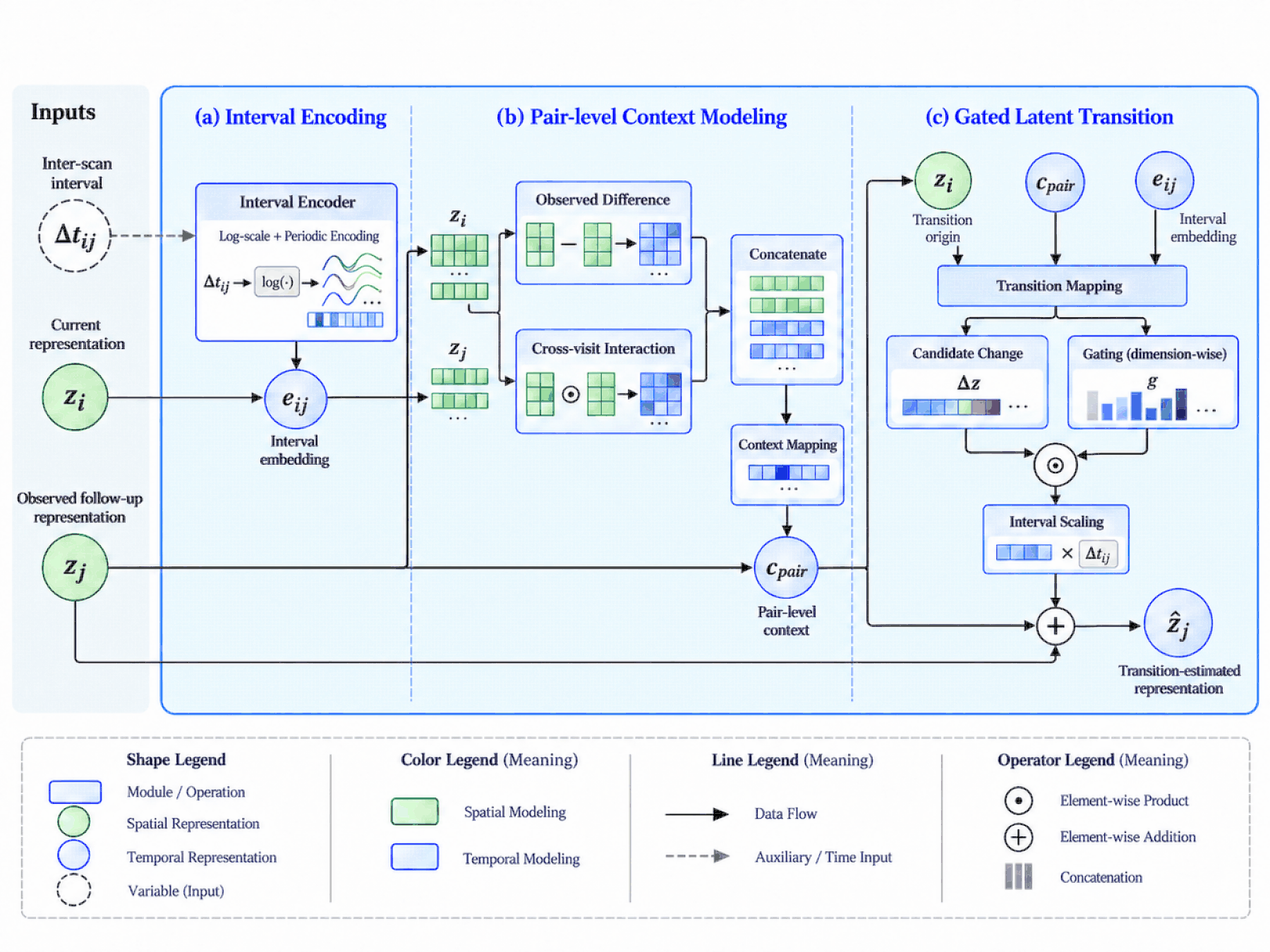}
    \caption{
\textbf{Time-conditioned latent transition.}
The inter-scan interval $\Delta t_{ij}$ is first log-scaled, periodically
encoded, and mapped to the interval embedding $\mathbf{e}_{ij}$. The observed
visit representations $\mathbf{z}_{i}$ and $\mathbf{z}_{j}$ are then used to
compute their latent difference and adaptively weighted cross-visit
interaction, which together form the pair-level context
$\mathbf{c}_{\mathrm{pair}}$. The earlier representation, pair-level context,
and interval embedding are subsequently used to estimate an interval-scaled
gated latent change, producing the transition-estimated representation
$\hat{\mathbf{z}}_{j}$.
}
    \label{fig:latent_transition}
\end{figure*}

The spatial module provides representations of the two observed MRI scans.
To characterize their relationship over irregular follow-up intervals, the
time-conditioned latent transition combines inter-scan interval encoding,
pair-level context modeling, and gated transition estimation
(Figure~\ref{fig:latent_transition}).

\subsubsection{Inter-scan Interval Encoding}

To condition transition modeling on follow-up timing, the inter-scan
interval is log-scaled and mapped to a temporal embedding:

\begin{equation}
\bar{\Delta t}_{ij}
=
\log
\left(
1+\frac{\Delta t_{ij}}{\tau}
\right),
\qquad
\mathbf{e}_{ij}
=
f_{\mathrm{time}}
\left[
\gamma(\bar{\Delta t}_{ij})
\right],
\label{eq:time_embedding}
\end{equation}
where
$\mathbf{e}_{ij}\in\mathbb{R}^{E}$,
$\tau>0$ is the temporal scaling constant,
$\gamma(\cdot)$ combines the log-scaled interval with periodic temporal
features, and $f_{\mathrm{time}}(\cdot)$ is a multilayer perceptron.
The resulting $\mathbf{e}_{ij}$ conditions the subsequent latent transition.


\subsubsection{Pair-level Context Modeling}

Pair-level context summarizes the observed visit representations, their
change, and their cross-visit feature interaction.

The observed latent change is defined as

\begin{equation}
\Delta\mathbf{z}^{\mathrm{obs}}
=
\mathbf{z}_{j}-\mathbf{z}_{i}.
\label{eq:observed_change}
\end{equation}
where
$\Delta\mathbf{z}^{\mathrm{obs}}\in\mathbb{R}^{Z}$.

This difference captures additive change but not interactions between
corresponding latent dimensions. We therefore compute an adaptively weighted
cross-visit interaction:

\begin{equation}
\boldsymbol{\alpha}
=
\sigma
\left(
\mathbf{W}_{\alpha}
\left[
\mathbf{z}_{i};
\mathbf{z}_{j};
\Delta\mathbf{z}^{\mathrm{obs}}
\right]
+
\mathbf{b}_{\alpha}
\right),
\qquad
\mathbf{h}_{\mathrm{int}}
=
\boldsymbol{\alpha}
\odot
\left(
\mathbf{z}_{i}
\odot
\mathbf{z}_{j}
\right).
\label{eq:adaptive_hadamard}
\end{equation}
where
$\boldsymbol{\alpha}\in(0,1)^{Z}$,
$\mathbf{h}_{\mathrm{int}}\in\mathbb{R}^{Z}$,
$\mathbf{W}_{\alpha}\in\mathbb{R}^{Z\times3Z}$, and
$\mathbf{b}_{\alpha}\in\mathbb{R}^{Z}$ are learnable parameters.

The pair-level context then combines the observed visit representations with
the observed latent change and weighted cross-visit interaction:

\begin{equation}
\mathbf{c}_{\mathrm{pair}}
=
f_{\mathrm{ctx}}
\left[
\mathbf{z}_{i};
\mathbf{z}_{j};
\Delta\mathbf{z}^{\mathrm{obs}};
\mathbf{h}_{\mathrm{int}}
\right].
\label{eq:pair_context}
\end{equation}
where
$\mathbf{c}_{\mathrm{pair}}\in\mathbb{R}^{C}$ and
$f_{\mathrm{ctx}}(\cdot)$ is a multilayer perceptron.


\subsubsection{Gated Latent Transition}

The earlier representation, pair-level context, and temporal embedding
are combined to estimate a gated transition over the inter-scan interval.

A transition representation
$\mathbf{h}_{\mathrm{dyn}}
=
f_{\mathrm{dyn}}
[\mathbf{z}_{i};
\mathbf{c}_{\mathrm{pair}};
\mathbf{e}_{ij}]$
is first computed, where
$\mathbf{h}_{\mathrm{dyn}}\in\mathbb{R}^{H}$ and
$f_{\mathrm{dyn}}(\cdot)$ denotes the transition mapping.

The candidate latent change and its element-wise gate are then computed as

\begin{equation}
\Delta\mathbf{z}
=
\mathbf{W}_{\Delta}\mathbf{h}_{\mathrm{dyn}},
\qquad
\mathbf{g}
=
\sigma
\left(
\mathbf{W}_{g}\mathbf{h}_{\mathrm{dyn}}
\right),
\label{eq:transition_gate}
\end{equation}
where
$\Delta\mathbf{z}\in\mathbb{R}^{Z}$,
$\mathbf{g}\in(0,1)^{Z}$, and
$\mathbf{W}_{\Delta},\mathbf{W}_{g}
\in\mathbb{R}^{Z\times H}$.

The transition-estimated follow-up representation is then computed as

\begin{equation}
\hat{\mathbf{z}}_{j}
=
\mathbf{z}_{i}
+
\bar{\Delta t}_{ij}
\left(
\mathbf{g}\odot\Delta\mathbf{z}
\right).
\label{eq:latent_transition}
\end{equation}
where
$\hat{\mathbf{z}}_{j}\in\mathbb{R}^{Z}$.

Here, $\mathbf{z}_{i}$ provides the transition origin,
$\mathbf{g}$ modulates the contribution of each latent dimension to the update,
and $\bar{\Delta t}_{ij}$ scales the estimated change according to the
inter-scan interval. The resulting $\hat{\mathbf{z}}_{j}$ provides a
transition-estimated follow-up representation that complements the directly
observed follow-up representation. Because the pair-level context is
constructed from both observed visits, $\hat{\mathbf{z}}_{j}$ is interpreted
as a pair-conditioned representation of interval-dependent change rather than
a future-state prediction derived from $\mathbf{z}_{i}$ alone.


\subsection{Observed--Transition Fusion}
\label{sec:fusion}

\begin{figure*}[t]
    \centering
    \includegraphics[width=1.0\linewidth]
    {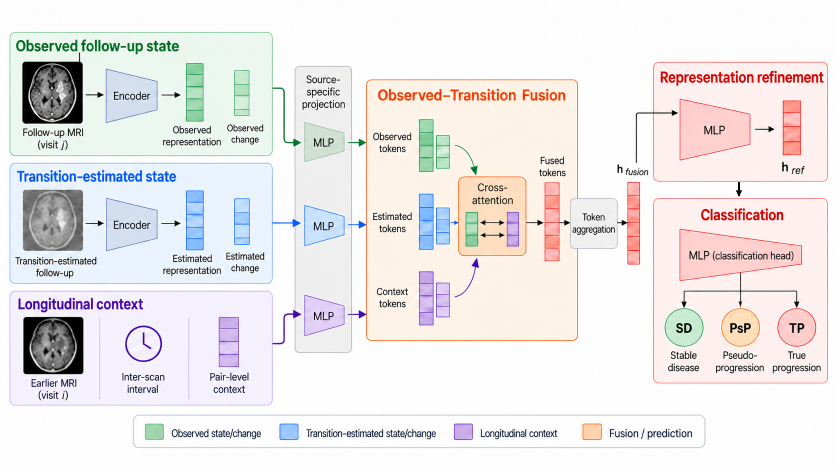}
    \caption{
\textbf{Observed--transition fusion.}
The fusion module integrates the earlier representation
$\mathbf{z}_{i}$, observed follow-up representation
$\mathbf{z}_{j}$, transition-estimated representation
$\hat{\mathbf{z}}_{j}$, observed and estimated latent changes,
pair-level context $\mathbf{c}_{\mathrm{pair}}$, and interval embedding
$\mathbf{e}_{ij}$. These longitudinal sources are first mapped into a common
representation space and integrated through multi-source cross-attention.
The resulting information is aggregated into the fused representation
$\mathbf{h}_{\mathrm{fusion}}$, which is refined and used for three-class
classification of SD, PsP, and TP.
}
    \label{fig:longitudinal_fusion}
\end{figure*}

The directly observed representation $\mathbf{z}_{j}\in\mathbb{R}^{Z}$
and transition-estimated representation
$\hat{\mathbf{z}}_{j}\in\mathbb{R}^{Z}$ provide complementary information
about the follow-up state. The fusion module integrates them with cross-visit
information for classification (Figure~\ref{fig:longitudinal_fusion}).

The transition-estimated change is defined as

\begin{equation}
\Delta\mathbf{z}^{\mathrm{est}}
=
\hat{\mathbf{z}}_{j}-\mathbf{z}_{i}.
\label{eq:estimated_change}
\end{equation}
where
$\Delta\mathbf{z}^{\mathrm{est}}\in\mathbb{R}^{Z}$.

To preserve both state and change information, the observed and
transition-estimated visit representations, their corresponding latent
changes, pair-level context, and temporal embedding are projected into a common
representation space:

\begin{equation}
\mathbf{S}_{0}
=
\left[
\mathcal{P}_{1}(\mathbf{z}_{i});
\mathcal{P}_{2}(\mathbf{z}_{j});
\mathcal{P}_{3}(\hat{\mathbf{z}}_{j});
\mathcal{P}_{4}(\Delta\mathbf{z}^{\mathrm{obs}});
\mathcal{P}_{5}(\Delta\mathbf{z}^{\mathrm{est}});
\mathcal{P}_{6}(\mathbf{c}_{\mathrm{pair}});
\mathcal{P}_{7}(\mathbf{e}_{ij})
\right],
\label{eq:pair_representation}
\end{equation}
where
$\mathbf{S}_{0}\in\mathbb{R}^{N_S\times Z}$,
$\mathcal{P}_{k}(\cdot)$ denotes a learnable source-specific projection,
and $N_S$ denotes the number of projected source tokens.

The projected sources are then processed by the learnable longitudinal fusion
operator $\mathcal{F}_{\mathrm{fusion}}$ and aggregated into a unified
representation:

\begin{equation}
\mathbf{h}_{\mathrm{fusion}}
=
\operatorname{Agg}
\left[
\mathcal{F}_{\mathrm{fusion}}
\left(
\mathbf{S}_{0}
\right)
\right],
\label{eq:fusion}
\end{equation}
where
$\mathbf{h}_{\mathrm{fusion}}\in\mathbb{R}^{F}$ and
$\operatorname{Agg}(\cdot)$ denotes the fusion aggregation operation.

Consistent with the prediction head in
Figure~\ref{fig:longitudinal_fusion}, the fused representation is further
refined by an MLP
$f_{\mathrm{ref}}:\mathbb{R}^{F}\rightarrow\mathbb{R}^{R}$, producing
$\mathbf{h}_{\mathrm{ref}}\in\mathbb{R}^{R}$. The refined representation is
then passed to a learnable classification head
$\operatorname{Cls}:\mathbb{R}^{R}\rightarrow\mathbb{R}^{K}$, followed by
softmax, to obtain the probability vector
$\mathbf{p}_{ij}\in[0,1]^K$.


\subsection{Training Objectives}
\label{sec:training_objectives}

The prediction output is optimized using a class-weighted classification
objective together with transition-specific regularization. Because LUMIERE
pretraining and Burdenko downstream adaptation use different label spaces,
the class-weighted cross-entropy loss is defined as

\begin{equation}
\mathcal{L}_{\mathrm{cls}}
=
-\sum_{k=1}^{K}
w_k y_{ij,k}\log p_{ij,k},
\label{eq:classification_loss}
\end{equation}
where $K$ denotes the number of classes at the corresponding training stage,
with $K=7$ for LUMIERE pretraining and $K=3$ for Burdenko downstream
adaptation. Here, $y_{ij,k}$ is the target indicator for class $k$ and $w_k$
is the corresponding class weight.

Because the transition-estimated representation is intended to remain
consistent with the observed follow-up representation, we further constrain
their latent-space discrepancy using

\begin{equation}
\mathcal{L}_{\mathrm{tr}}
=
\left\|
\hat{\mathbf{z}}_{j}
-
\mathbf{z}_{j}
\right\|_2^2,
\qquad
\mathcal{L}_{\mathrm{dir}}
=
1-
\cos
\left(
\hat{\mathbf{z}}_{j},
\mathbf{z}_{j}
\right),
\label{eq:transition_losses}
\end{equation}
where $\cos(\cdot,\cdot)$ denotes cosine similarity.
The transition loss $\mathcal{L}_{\mathrm{tr}}$ constrains the magnitude of
the discrepancy between the transition-estimated and observed follow-up
representations, whereas the latent directional consistency loss
$\mathcal{L}_{\mathrm{dir}}$ encourages directional alignment between the two
follow-up representations in latent space.

The complete objectives for longitudinal pretraining and downstream adaptation
are defined as

\begin{align}
\mathcal{L}_{\mathrm{pre}}
&=
\mathcal{L}_{\mathrm{cls}}
+
\lambda_{\mathrm{tr}}\mathcal{L}_{\mathrm{tr}}
+
\lambda_{\mathrm{dir}}\mathcal{L}_{\mathrm{dir}},
\\
\mathcal{L}_{\mathrm{down}}
&=
\mathcal{L}_{\mathrm{cls}}
+
\beta_{\mathrm{tr}}\mathcal{L}_{\mathrm{tr}}
+
\beta_{\mathrm{dir}}\mathcal{L}_{\mathrm{dir}},
\label{eq:overall_losses}
\end{align}
where $\mathcal{L}_{\mathrm{pre}}$ and
$\mathcal{L}_{\mathrm{down}}$ denote the objectives used for LUMIERE
pretraining and Burdenko downstream adaptation, respectively.
$\lambda_{\mathrm{tr}}$ and $\lambda_{\mathrm{dir}}$ control the contributions
of the transition and latent directional consistency losses during pretraining,
whereas $\beta_{\mathrm{tr}}$ and $\beta_{\mathrm{dir}}$ control their
contributions during downstream adaptation.


\section{Experimental Setup}
\label{sec:experimental_setup}


\subsection{Datasets and Data Preparation}
\label{sec:cohort_results}

We used BraTS2024~\citep{de20242024} for lesion-prior generation,
LUMIERE~\citep{suter2022lumiere} for longitudinal pretraining, and the
Burdenko Glioblastoma Progression Dataset~\citep{zolotova2023burdenko}
for downstream three-class clinical-state classification
(Table~\ref{tab:cohort_summary}).

\begin{table*}[t]
\centering
\footnotesize
\setlength{\tabcolsep}{4pt}
\renewcommand{\arraystretch}{1.18}

\caption{
Characteristics and study roles of the three datasets.
Age (mean $\pm$ standard deviation) and sex are summarized per patient,
whereas clinical-label counts are reported per longitudinal pair.
Statistics for LUMIERE and Burdenko correspond to the retained cohorts after
quality control and longitudinal sample construction.
Abbreviations: M/F, male/female; WT, whole tumor; TC, tumor core;
ET, enhancing tumor; RC, resection cavity; PD, progressive disease;
SD, stable disease; PR, partial response; CR, complete response;
PsP, pseudoprogression; TP, true progression; N/A, not applicable.
}
\label{tab:cohort_summary}

\begin{adjustbox}{max width=\textwidth}
\begin{tabular}{
p{2.3cm}
p{2.8cm}
p{3.8cm}
p{3.6cm}
}
\toprule

\textbf{Characteristic}
&
\textbf{BraTS2024}
&
\textbf{LUMIERE}
&
\textbf{Burdenko}
\\

\midrule

Study role
&
Lesion-prior generation
&
Longitudinal pretraining
&
Downstream classification
\\

Patients
&
613
&
87
&
98
\\

Age (years)
&
52.9 $\pm$ 14.1
&
62.1 $\pm$ 10.3
&
54.5 $\pm$ 12.2
\\

Sex (M/F)
&
355/258
&
44/43
&
50/48
\\

MRI scans
&
1,350
&
591
&
279
\\

Mean scans/patient
&
2.20
&
6.79
&
2.85
\\

Longitudinal pairs
&
N/A
&
499
&
181
\\

Clinical labels (pairs)
&
N/A
&
PD, 233 (46.7\%); SD, 93 (18.6\%);
PR, 20 (4.0\%); CR, 25 (5.0\%);
post-operative, 99 (19.8\%);
pre-operative, 1 (0.2\%);
unknown, 28 (5.6\%)
&
SD, 39 (21.5\%);
PsP, 31 (17.1\%);
TP, 111 (61.3\%)
\\

Segmentation
&
WT, TC, ET, and RC
&
Automated tumor masks
&
N/A
\\

Temporal information
&
N/A
&
Longitudinal scan timing
&
Longitudinal follow-up timing
\\

\bottomrule
\end{tabular}
\end{adjustbox}
\end{table*}

We used all 1,350 annotated examinations from 613 patients in the initial
BraTS2024 Adult Glioma Post-Treatment training release, each with MNI-space
T1n, T1c, T2w, T2-FLAIR, and a voxelwise reference segmentation. The
Additional Training Data (271 examinations from 118 patients) was not used to
train the segmentation ensemble; the 188-examination validation release was
excluded because it lacked publicly available reference segmentations.

Longitudinal MRI scans from LUMIERE and Burdenko underwent the same
preprocessing pipeline. Raw data were converted from DICOM to NIfTI and
screened for voxel spacing, image geometry, slice thickness, and image
clarity. Retained volumes were rigidly registered to the MNI152 template,
skull-stripped, and foreground z-score normalized. Each volume was resampled
to an isotropic voxel spacing of
$1.25\times1.25\times1.25$~mm$^3$ and adjusted to
$128\times128\times128$ voxels through foreground cropping, center padding, or
center cropping. These steps yield standardized 3D inputs rather than
native-resolution, full-field-of-view MRI, allowing consistent volumetric
modeling across examinations with different acquisition geometries.

Before spatial representation learning, Burdenko scans were associated with
abnormal-region probability maps from the BraTS2024-trained segmentation
ensemble, whereas LUMIERE scans used the available automated tumor masks.
These lesion priors served only as auxiliary structural guidance.

Longitudinal pairs were then constructed independently within each patient.
The LUMIERE release includes 91 patients and 638 study dates, of which 599
provide all four MRI sequences. Visits were ordered by encoded follow-up week
and acquisition phase and retained when contrast-enhanced T1-weighted,
native T1-weighted, T2-weighted, and FLAIR images were all available.
Consecutive eligible visits were paired, and the later visit's seven-category
annotation was used as the pretraining target.

\begin{figure}[t]
    \centering
    \includegraphics[width=1.0\linewidth]{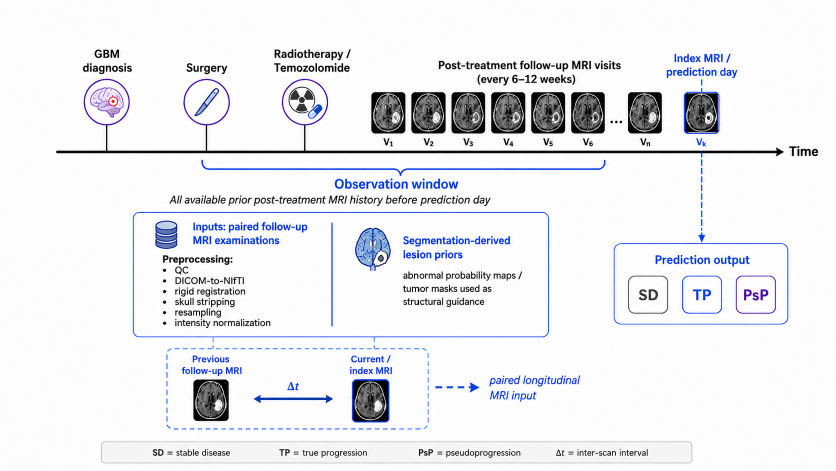}
    \caption{
    \textbf{Longitudinal study setup for post-treatment GBM clinical-state
    classification.}
    Eligible post-treatment MRI scans and their lesion priors form
    adjacent-visit pairs separated by the inter-scan interval $\Delta t$.
    Each pair is used to predict the patient's curated clinical-state label:
    stable disease (SD), pseudoprogression (PsP), or true progression (TP).
    }
    \label{fig:longitudinal_setup}
\end{figure}

From the 180-patient Burdenko release, we retained patients with curated SD,
PsP, or TP clinical-state labels. Follow-up eligibility began at the first
recorded follow-up date, with the topometric MRI date and then the
end-of-radiotherapy date used as fallback anchors. Only visits acquired on or
after the applicable anchor with contrast-enhanced T1-weighted and FLAIR
images were retained. Native T1-weighted and T2-weighted images were included
when available, with modality-presence indicators supplied to the model.
Patients with fewer than two eligible visits were excluded. The remaining
visits were ordered chronologically, and only adjacent visits were paired.
For each pair, the inter-scan interval was
calculated in weeks as
$\Delta t=(d_{t+1}-d_t)/7$, where $d_t$ and $d_{t+1}$ denote the acquisition
dates in days. Each pair inherited the patient's curated SD, PsP, or TP
clinical-state label; thus, eligible pairs from the same patient shared that
label. The resulting longitudinal pair construction and
downstream prediction setup are illustrated in
Figure~\ref{fig:longitudinal_setup}.

\subsection{Training and Implementation Details}

Model development proceeded in three stages using PyTorch. The segmentation
ensemble was trained on BraTS2024 and frozen; STRIDE was pretrained on
LUMIERE paired MRI scans and their inter-scan intervals, and its parameters
initialized the Burdenko model for downstream SD, PsP, and TP
classification. The optimization configurations are summarized in
Table~\ref{tab:training_configuration}. A weight decay of $1\times10^{-4}$ was
used throughout training, and gradients were clipped to a maximum $\ell_2$ norm
of 1.0.

\begin{table}[t]
\centering
\caption{
Training configurations for segmentation training, longitudinal pretraining,
and downstream clinical-state classification.
}
\label{tab:training_configuration}

\begin{adjustbox}{max width=\textwidth}
\begin{tabular}{lcccc}
\toprule
Stage
& Optimizer
& Learning rate
& Batch size
& Epochs
\\
\midrule

BraTS2024 segmentation
& AdamW
& $1\times10^{-3}$
& 2
& 200
\\

LUMIERE pretraining
& AdamW
& $1\times10^{-4}$
& 8
& 300
\\

Burdenko fine-tuning
& AdamW
& $8\times10^{-5}$
& 8
& 200
\\

\bottomrule
\end{tabular}
\end{adjustbox}
\end{table}

Data augmentation was applied only to training samples. Each spatial axis was
independently flipped with probability 0.5, with the same spatial
transformation applied jointly to both scans in each longitudinal pair, all MRI
channels, and their corresponding lesion masks. Intensity augmentation
included random scaling $a\sim\mathcal{U}(0.9,1.1)$, random shifting
$b\sim\mathcal{U}(-0.1,0.1)$, and additive Gaussian noise
$\boldsymbol{\epsilon}\sim\mathcal{N}(0,0.03^2)$.
Validation and test samples were not augmented.

Class imbalance was addressed through the class-weighted cross-entropy loss in
Eq.~\ref{eq:classification_loss}. Class weights were calculated from normalized
inverse class frequencies within each training partition and scaled to have a
mean of one. For the transition-related objectives, the pretraining
coefficients were set
to $\lambda_{\mathrm{tr}}=0.25$ and
$\lambda_{\mathrm{dir}}=0.10$, while the downstream coefficients were
$\beta_{\mathrm{tr}}=0.15$ and
$\beta_{\mathrm{dir}}=0.05$.

For downstream classification, model checkpoints were selected according to
the highest macro-averaged validation ROC--AUC. During Burdenko fine-tuning,
the first 10 epochs were excluded from checkpoint selection. The selected
checkpoint was evaluated on the held-out test fold, which was never used for
optimization or checkpoint selection. The random seed was fixed at 42 for
patient partitioning, data-loader initialization, and model training.

During downstream transfer, the pretrained backbone was frozen for the first
five epochs and subsequently optimized jointly with the temporal and
classification modules. Additional implementation settings included a stem
width of 12, 24 hidden channels in the multi-scale localization branches, a
drop-path probability of 0.15, an input-adaptation dropout of 0.08, and a
temporal-dynamics dropout of 0.10. The temporal scaling constant in
Eq.~\ref{eq:time_embedding} was set to $\tau=52$ weeks. Eight periodic
frequencies were used for temporal encoding and projected to a 64-dimensional
temporal embedding. The pair-context representation used an output dimension
of 192 and a hidden dimension of 384.


\subsection{Evaluation Protocol and Metrics}

For lesion segmentation, the labeled BraTS2024 examinations were assessed
using five examination-level folds supplied with the segmentation ensemble:
four for training and one for validation in each iteration, without an
independent test set. Among 484 patients with multiple examinations,
415 (85.7\%) had scans in multiple folds; thus, the reported segmentation
metrics are not patient-independent.

To evaluate downstream classification while preventing patient-level
information leakage, performance on Burdenko was assessed using patient-level
stratified ten-fold cross-validation. All MRI scans and longitudinal pairs
derived from the same patient were assigned to the same fold. In each
cross-validation iteration, one fold was reserved for final testing, one fold
was used for validation and model selection, and the remaining eight folds
were used for training.

Longitudinal modeling strategies were compared using identical patient-level
partitions and evaluation protocols. The evaluated strategies included direct
feature fusion, Siamese feature comparison, LSTM-based temporal aggregation,
Temporal Transformer modeling, and STRIDE. Whenever applicable, the same
visit-level spatial encoder, lesion-prior guidance, data augmentation, and
optimization settings were maintained so that the comparison primarily
reflected differences in longitudinal modeling strategy.

For three-class post-treatment clinical-state classification, macro-averaged
one-vs-rest ROC--AUC was used as the primary evaluation metric. Accuracy and
macro F1-score were additionally reported. For lesion segmentation, Dice
coefficient and Hausdorff distance at the 95th percentile (HD95) were used to
evaluate volumetric overlap and boundary agreement, respectively. All
classification metrics were calculated on held-out test samples within the
patient-level cross-validation procedure. Classification results are reported
as mean $\pm$ standard deviation across the ten held-out test folds. A
pooled out-of-fold macro ROC--AUC was also computed once after aggregating
the predictions from all ten test folds.

For comparisons with longitudinal baselines, the ten test folds were matched
by patient-level partition. Within each fold, the baseline macro ROC--AUC was
subtracted from that of STRIDE. The mean difference and its two-sided 95\%
confidence interval were calculated from these ten paired differences using
the Student $t$ distribution with nine degrees of freedom. Two-sided paired
$t$ tests were performed on the same differences, and the resulting four
$p$ values were adjusted using Holm's method. Because the training sets
overlapped across folds, these comparisons were interpreted as exploratory
within-cohort analyses.

\section{Results}
\label{sec:results}


\subsection{Segmentation Performance and Effect of Lesion-Prior Guidance}
\label{sec:rq1}

\begin{figure*}[t]
\centering
\includegraphics[width=0.95\linewidth]{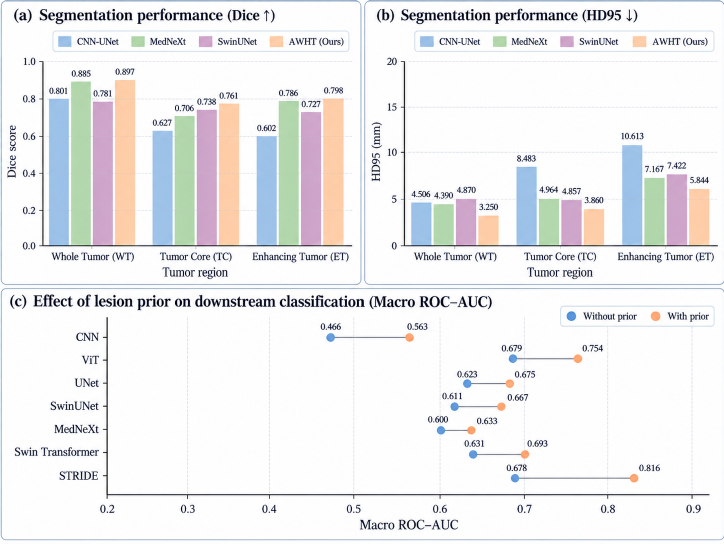}

\caption{
Segmentation performance and downstream effect of lesion-prior guidance.
(a) Dice scores for whole tumor (WT), tumor core (TC), and enhancing tumor
(ET) on BraTS2024.
(b) HD95 values for the corresponding tumor regions, where lower values
indicate better boundary agreement.
(c) Macro ROC--AUC comparison with and without segmentation-derived lesion
priors across different spatial backbones. Error bars denote standard
deviation across the ten held-out test folds where applicable.
}
\label{fig:lesion_prior_results}
\end{figure*}

To evaluate lesion localization, the AWHT-based segmentation network was
compared with representative segmentation backbones on BraTS2024.
As shown in Fig.~\ref{fig:lesion_prior_results}(a,b), AWHT achieved the
highest Dice scores and the lowest HD95 values across whole tumor (WT),
tumor core (TC), and enhancing tumor (ET). Specifically, Dice scores reached
0.897, 0.761, and 0.798 for WT, TC, and ET, respectively, while the
corresponding HD95 values were 3.250, 3.860, and 5.844 mm.
Complete numerical comparisons are provided in
Table~\ref{tab:app_brats_segmentation}.

We next examined whether these segmentation-derived priors translated into
improved clinical-state classification. As shown in
Fig.~\ref{fig:lesion_prior_results}(c), lesion-prior guidance increased
macro ROC--AUC across all evaluated architectures. Averaged across models,
macro ROC--AUC increased from 0.613 to 0.686. Average accuracy increased
from 0.645 to 0.714, while macro F1 increased from 0.609 to 0.668.
Complete results for all three metrics are reported in
Table~\ref{tab:app_mask_guidance}.


\subsection{Longitudinal Pretraining and Benefits of Longitudinal Input}
\label{sec:rq2}

\begin{figure*}[t]
\centering
\includegraphics[width=0.95\linewidth]{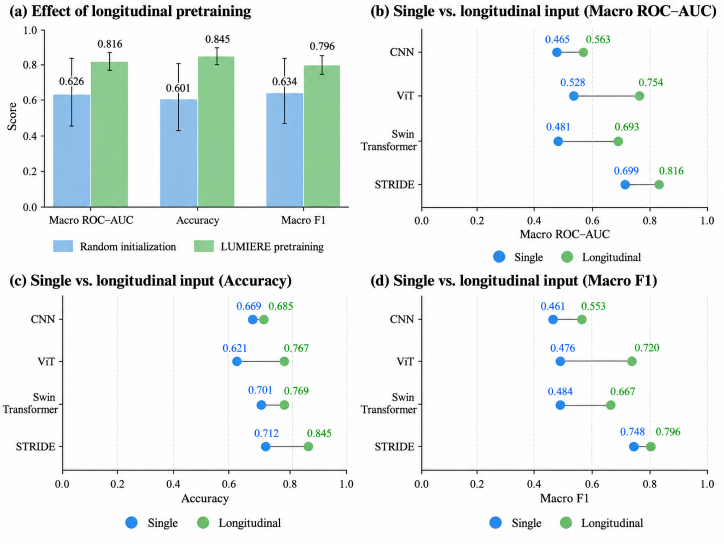}

\caption{
Benefits of longitudinal learning for downstream post-treatment GBM
clinical-state classification.
(a) Comparison between random initialization and LUMIERE longitudinal
pretraining.
(b--d) Comparison between single-timepoint and paired longitudinal inputs
for macro ROC--AUC, accuracy, and macro F1, respectively.
Error bars denote standard deviation across the ten held-out test folds.
}
\label{fig:longitudinal_benefits}
\end{figure*}

STRIDE was pretrained on the 499 longitudinal LUMIERE pairs and subsequently
adapted to Burdenko, with random initialization used as the reference under
identical downstream training settings. As shown in
Fig.~\ref{fig:longitudinal_benefits}(a), longitudinal pretraining increased
macro ROC--AUC from 0.626 to 0.816, accuracy from 0.601 to 0.845, and macro F1
from 0.634 to 0.796. Complete mean $\pm$ standard deviation results are
provided in Table~\ref{tab:app_pretraining}.

We then compared single-timepoint and paired longitudinal configurations
across multiple spatial backbones under the same evaluation setting.
As shown in Fig.~\ref{fig:longitudinal_benefits}(b--d), longitudinal input
improved performance across all evaluated architectures. Averaged across
models, macro ROC--AUC increased from 0.543 to 0.706, accuracy from 0.676
to 0.767, and macro F1 from 0.542 to 0.684. Complete numerical results are
reported in Table~\ref{tab:app_single_longitudinal}.


\subsection{Comparison with Longitudinal Modeling Strategies}
\label{sec:longitudinal_baselines}

\begin{figure*}[t]
\centering
\includegraphics[width=0.95\linewidth]{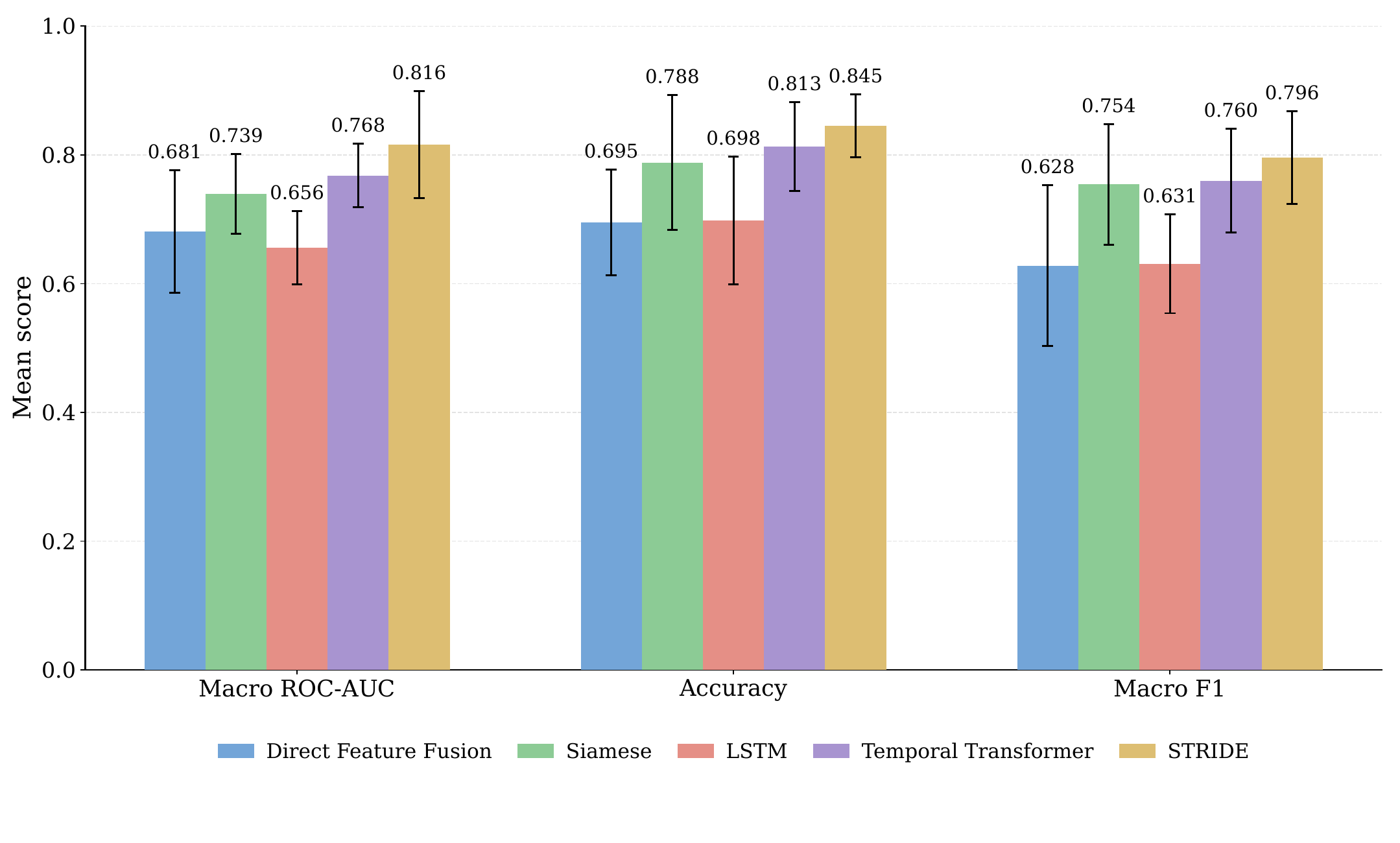}

\caption{
Comparison of longitudinal modeling strategies for three-class
post-treatment GBM clinical-state classification. Performance is reported
using macro ROC--AUC, accuracy, and macro F1. Error bars denote standard
deviation across the ten held-out test folds.
}
\label{fig:longitudinal_baselines}
\end{figure*}

STRIDE was compared with direct feature fusion, Siamese feature comparison,
recurrent temporal aggregation, and attention-based temporal modeling.

As shown in Fig.~\ref{fig:longitudinal_baselines}, STRIDE achieved the highest
mean performance across all three evaluation metrics, reaching a fold-wise
mean macro ROC--AUC of 0.816 $\pm$ 0.083, accuracy of 0.845 $\pm$ 0.049,
and macro F1 of 0.796 $\pm$ 0.072. Complete numerical results are provided
in Table~\ref{tab:app_longitudinal_baselines}.

The Temporal Transformer was the closest competitor, with a fold-wise mean
macro ROC--AUC of 0.768, accuracy of 0.813, and macro F1 of 0.760
(Table~\ref{tab:app_longitudinal_baselines}). The mean within-fold
macro ROC--AUC difference between STRIDE and this baseline was 0.048.
Paired comparisons with all four baselines are reported in
Table~\ref{tab:app_paired_auc}.

Pooled out-of-fold predictions yielded a macro ROC--AUC of 0.85
(Fig.~\ref{fig:app_classwise_performance}), compared with
the mean fold-wise macro ROC--AUC of 0.816 reported above. The same figure
shows class-wise precision--recall curves and a normalized confusion matrix
for SD, PsP, and TP.

\subsection{Component Analysis of STRIDE}
\label{sec:ablation}
\begin{figure*}[t]
\centering
\includegraphics[width=0.95\linewidth]{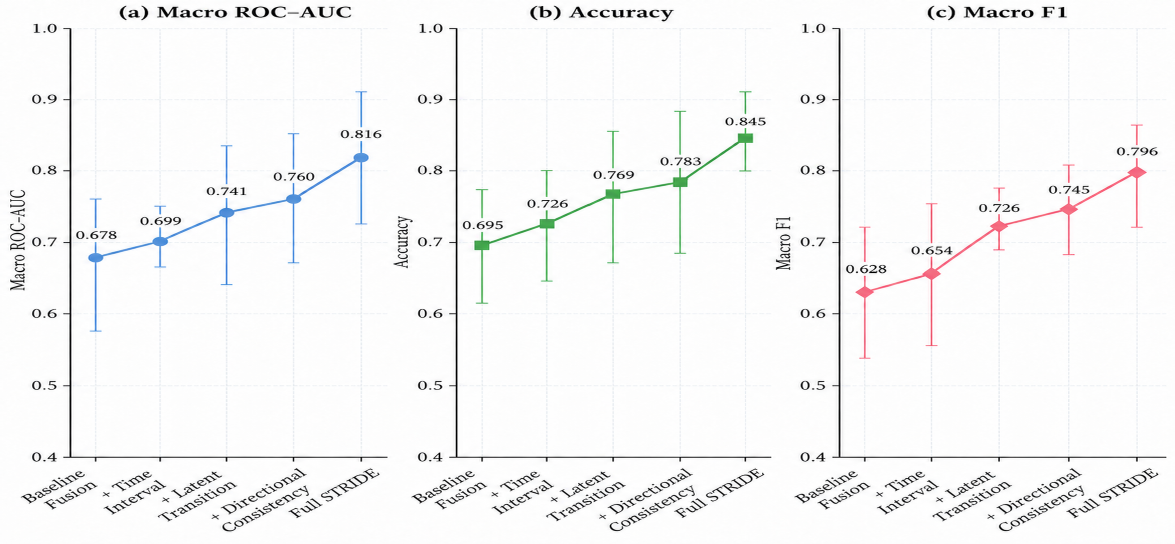}

\caption{
Cumulative component analysis of STRIDE.
Performance changes are shown for
(a) macro ROC--AUC,
(b) accuracy, and
(c) macro F1 as interval conditioning, latent transition modeling,
directional consistency, and observed--transition fusion are progressively
introduced. Error bars denote standard deviation across the ten held-out
test folds.
}
\label{fig:stride_ablation}
\end{figure*}
A cumulative ablation progressively introduced interval conditioning,
gated latent transition modeling, latent directional consistency
regularization, and observed--transition fusion.

As shown in Fig.~\ref{fig:stride_ablation}, performance improved
progressively as additional components were introduced. Interval conditioning
increased macro ROC--AUC from 0.678 to 0.699. Adding the gated latent
transition further increased ROC--AUC to 0.741 and macro F1 from 0.654 to
0.726. Adding latent directional consistency increased macro ROC--AUC to
0.760, and the complete configuration with observed--transition fusion
achieved the highest performance, reaching 0.816 macro ROC--AUC,
0.845 accuracy, and 0.796 macro F1. The complete mean $\pm$ standard deviation results and exact component
configuration for each ablation setting are provided in
Table~\ref{tab:app_temporal_ablation}.

\begin{figure*}[t]
\centering
\includegraphics[width=0.95\linewidth]
{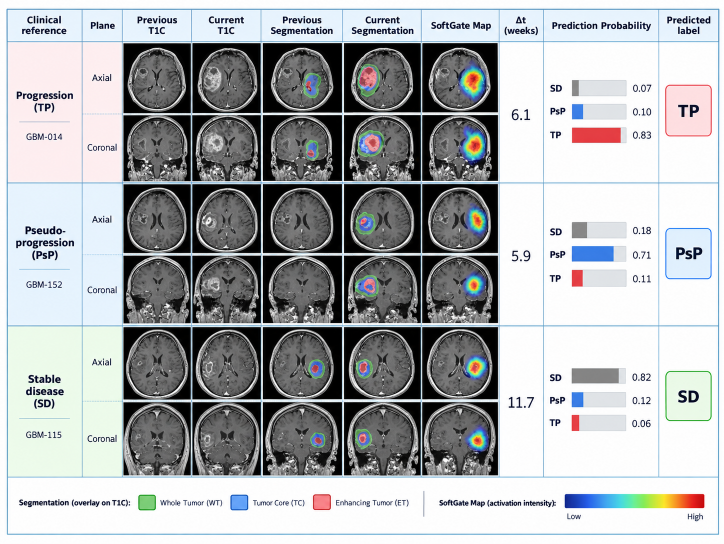}

\caption{
Qualitative visualization of representative longitudinal GBM cases from
true progression (TP), pseudoprogression (PsP), and stable disease (SD).
For each case, previous and current contrast-enhanced T1-weighted MRI scans
are shown together with segmentation-derived lesion priors, SoftGate
activation maps, the inter-scan interval, class probabilities, and the final
predicted label.
}
\label{fig:qualitative_results}
\end{figure*}
\subsection{Qualitative Analysis of Lesion-aware Longitudinal Representations}
\label{sec:qualitative_results}

Figure~\ref{fig:qualitative_results} presents lesion priors and SoftGate
activation maps for representative TP, PsP, and SD cases. Across the representative cases, the lesion priors identify tumor-related
regions in both examinations, while SoftGate emphasizes these regions without
entirely suppressing surrounding tissue. The spatial emphasis also changes
between longitudinal visits in accordance with changes in visible lesion
distribution rather than remaining fixed across examinations.

\section{Discussion}
\label{sec:discussion}

The improvements associated with lesion-prior guidance and paired MRI input
suggest complementary contributions from spatial representation and
longitudinal information. Previous work has examined GBM follow-up MRI at
individual timepoints ~\citep{guo2025timepoint} and longitudinal MRI in
high-grade glioma surveillance ~\citep{siakallis2021longitudinal}. In the
present cohort, combining scan-specific spatial guidance with longitudinal
modeling helped distinguish SD, PsP, and TP. The qualitative findings are
consistent with the intended role of SoftGate: emphasizing lesion-related
regions while retaining surrounding context.

Variable follow-up timing is a broader challenge in irregularly sampled
longitudinal modeling ~\citep{chen2018neural,rubanova2019latent}. The
cumulative component analysis is consistent with improved performance when
interval conditioning and estimated follow-up information are added
alongside the observed follow-up representation. Because the components
were added in sequence, their independent effects cannot be determined
from this ablation. STRIDE also achieved a higher mean fold-wise macro
ROC--AUC than the closest Temporal Transformer baseline, with paired
comparisons reported in Table~\ref{tab:app_paired_auc}. These
comparisons reflect performance within the Burdenko cohort.

The downstream evaluation used patient-level stratified ten-fold
cross-validation within a single retrospective Burdenko cohort. Keeping all
scans and pairs from each patient in one fold limited patient-level leakage,
but all eligible pairs from that patient shared a curated clinical-state
label. The resulting pair-level predictions do not establish visit-specific
diagnostic accuracy. Moreover, the BraTS2024 segmentation evaluation used
examination-level folds with patient overlap between training and validation,
which may have led to optimistic segmentation performance estimates. The
quality of the segmentation network's generated priors on Burdenko was not
quantitatively assessed. Independent downstream validation and assessment
of lesion-prior quality on post-treatment scans are needed to establish
generalizability and clinical utility.

\section{Conclusion}
\label{sec:conclusion}

We presented STRIDE, a longitudinal Transformer framework for three-class
post-treatment GBM clinical-state classification from paired MRI scans and
their inter-scan interval. STRIDE combines lesion-prior-guided spatial
representation, time-conditioned latent transition, and observed--transition
fusion in a unified framework. On Burdenko, it achieved the highest mean
performance among the evaluated longitudinal strategies. These results
support further evaluation of interval-conditioned disease representations
on independent post-treatment GBM cohorts.

\section*{Data and code availability}
The BraTS2024, LUMIERE, and Burdenko Glioblastoma Progression datasets are
publicly available through the BraTS challenge platform, the corresponding
LUMIERE data repository, and The Cancer Imaging Archive (TCIA), respectively.

The code used to implement the proposed STRIDE framework will be made publicly
available on GitHub upon publication at
\url{https://github.com/AdrianGuo2484/STRIDE}.

\section*{Acknowledgments}
Computational resources were provided by the Ohio Supercomputer Center (OSC).

\section*{Declaration of competing interest}
The authors declare that they have no known competing financial interests or
personal relationships that could have appeared to influence the work reported
in this paper.

\section*{CRediT authorship contribution statement}
W.G.: Conceptualization, Methodology, Software, Investigation, Formal analysis,
Writing -- original draft.
C.Y.: Methodology. C.Y. also provided technical guidance and assisted with
interpretation of experimental results.
W.C.: Supervision, Writing -- review \& editing.
G.M.: Methodology, Supervision, Writing -- review \& editing.
P.Z.: Methodology, Writing -- review \& editing.
P.G.: Methodology, Writing -- review \& editing. P.G. also contributed to
clinical interpretation of the results.

\section*{Declaration of generative AI and AI-assisted technologies in the
manuscript preparation process}
During the preparation of this work, the authors used OpenAI ChatGPT to assist
with language editing. The authors reviewed and
edited the resulting content and take full responsibility for the content of the published article.


\FloatBarrier
\clearpage
\appendix

\section{Additional Quantitative Results}
\label{sec:appendix_quantitative_results}

This appendix provides the complete numerical results
corresponding to the quantitative comparisons presented in the main
manuscript. Tables A.1--A.7 report the detailed performance values supporting
the segmentation evaluation, lesion-prior guidance analysis, longitudinal
pretraining experiment, single-versus-longitudinal comparison, longitudinal
modeling strategy comparison, STRIDE component analysis, and paired
comparison with the evaluated baselines, respectively. Figure A.1 additionally
presents class-wise discrimination and prediction patterns of the final
STRIDE model.


\setcounter{table}{0}
\renewcommand{\thetable}{A.\arabic{table}}

\setcounter{figure}{0}
\renewcommand{\thefigure}{A.\arabic{figure}}

\FloatBarrier


\begin{table*}[!htbp]
\centering
\small

\caption{
Complete segmentation performance comparison on BraTS2024.
Dice ($\uparrow$) and HD95 ($\downarrow$, mm) are reported for whole tumor
(WT), tumor core (TC), and enhancing tumor (ET).
}
\label{tab:app_brats_segmentation}

\begin{adjustbox}{max width=\textwidth}
\begin{tabular}{lcccccc}
\toprule

Backbone
& \multicolumn{3}{c}{Dice $\uparrow$}
& \multicolumn{3}{c}{HD95 $\downarrow$}
\\

\cmidrule(lr){2-4}
\cmidrule(lr){5-7}

& WT & TC & ET & WT & TC & ET
\\

\midrule

CNN-UNet
& 0.801
& 0.627
& 0.602
& 4.506
& 8.483
& 10.613
\\

MedNeXt
& 0.885
& 0.706
& 0.786
& 4.390
& 4.964
& 7.167
\\

SwinUNet
& 0.781
& 0.738
& 0.727
& 4.870
& 4.857
& 7.422
\\

AWHT Encoder (Ours)
& \textbf{0.897}
& \textbf{0.761}
& \textbf{0.798}
& \textbf{3.250}
& \textbf{3.860}
& \textbf{5.844}
\\

\bottomrule
\end{tabular}
\end{adjustbox}
\end{table*}


\begin{table*}[!htbp]
\centering
\footnotesize
\setlength{\tabcolsep}{4pt}

\caption{
Complete comparison of segmentation-derived lesion-prior guidance for
three-class post-treatment clinical-state classification.
Macro-averaged one-vs-rest ROC--AUC, accuracy, and macro F1 are reported as
mean $\pm$ standard deviation across the ten held-out test folds.
The Average row reports performance averaged across the evaluated
architectures.
}
\label{tab:app_mask_guidance}

\begin{adjustbox}{max width=\textwidth}
\begin{tabular}{lcccccc}
\toprule

& \multicolumn{2}{c}{Macro ROC--AUC $\uparrow$}
& \multicolumn{2}{c}{Accuracy $\uparrow$}
& \multicolumn{2}{c}{Macro F1 $\uparrow$}
\\

\cmidrule(lr){2-3}
\cmidrule(lr){4-5}
\cmidrule(lr){6-7}

Backbone
& w/o Prior
& w/ Prior
& w/o Prior
& w/ Prior
& w/o Prior
& w/ Prior
\\

\midrule

CNN
& 0.466 $\pm$ 0.197
& 0.563 $\pm$ 0.155
& 0.615 $\pm$ 0.167
& 0.685 $\pm$ 0.136
& 0.590 $\pm$ 0.141
& 0.553 $\pm$ 0.104
\\

ViT
& 0.679 $\pm$ 0.138
& 0.754 $\pm$ 0.075
& 0.634 $\pm$ 0.044
& 0.767 $\pm$ 0.096
& 0.592 $\pm$ 0.068
& 0.720 $\pm$ 0.142
\\

UNet
& 0.623 $\pm$ 0.071
& 0.675 $\pm$ 0.068
& 0.638 $\pm$ 0.102
& 0.615 $\pm$ 0.034
& 0.637 $\pm$ 0.141
& 0.654 $\pm$ 0.155
\\

SwinUNet
& 0.611 $\pm$ 0.072
& 0.667 $\pm$ 0.049
& 0.654 $\pm$ 0.088
& 0.662 $\pm$ 0.064
& 0.612 $\pm$ 0.058
& 0.633 $\pm$ 0.095
\\

MedNeXt
& 0.600 $\pm$ 0.137
& 0.633 $\pm$ 0.094
& 0.646 $\pm$ 0.086
& 0.653 $\pm$ 0.050
& 0.626 $\pm$ 0.139
& 0.654 $\pm$ 0.184
\\

Swin Transformer
& 0.631 $\pm$ 0.051
& 0.693 $\pm$ 0.038
& 0.667 $\pm$ 0.037
& 0.769 $\pm$ 0.049
& 0.575 $\pm$ 0.076
& 0.667 $\pm$ 0.056
\\

\textbf{STRIDE}
& \textbf{0.678 $\pm$ 0.116}
& \textbf{0.816 $\pm$ 0.083}
& \textbf{0.659 $\pm$ 0.171}
& \textbf{0.845 $\pm$ 0.049}
& \textbf{0.634 $\pm$ 0.077}
& \textbf{0.796 $\pm$ 0.072}
\\

\midrule

\textbf{Average}
& 0.613
& \textbf{0.686}
& 0.645
& \textbf{0.714}
& 0.609
& \textbf{0.668}
\\

\bottomrule
\end{tabular}
\end{adjustbox}
\end{table*}


\begin{table}[!htbp]
\centering
\small

\caption{
Effect of longitudinal pretraining on downstream three-class post-treatment
GBM clinical-state classification. Results are reported as mean $\pm$
standard deviation across the ten held-out test folds.
}
\label{tab:app_pretraining}

\begin{adjustbox}{max width=\textwidth}
\begin{tabular}{lccc}
\toprule

Initialization
& Macro ROC--AUC $\uparrow$
& Accuracy $\uparrow$
& Macro F1 $\uparrow$
\\

\midrule

Random initialization
& 0.626 $\pm$ 0.192
& 0.601 $\pm$ 0.221
& 0.634 $\pm$ 0.174
\\

LUMIERE pretraining
& \textbf{0.816 $\pm$ 0.083}
& \textbf{0.845 $\pm$ 0.049}
& \textbf{0.796 $\pm$ 0.072}
\\

\bottomrule
\end{tabular}
\end{adjustbox}
\end{table}


\begin{table*}[!htbp]
\centering
\footnotesize
\setlength{\tabcolsep}{3pt}

\caption{
Complete comparison between single-timepoint and longitudinal modeling for
three-class post-treatment classification. Macro-averaged one-vs-rest
ROC--AUC, accuracy, and macro F1 are reported as mean $\pm$ standard
deviation across the ten held-out test folds. The Average row reports
performance averaged across the evaluated architectures.
}
\label{tab:app_single_longitudinal}

\begin{adjustbox}{max width=\textwidth}
\begin{tabular}{lcccccc}
\toprule

& \multicolumn{2}{c}{Macro ROC--AUC $\uparrow$}
& \multicolumn{2}{c}{Accuracy $\uparrow$}
& \multicolumn{2}{c}{Macro F1 $\uparrow$}
\\

\cmidrule(lr){2-3}
\cmidrule(lr){4-5}
\cmidrule(lr){6-7}

Model
& Single
& Longitudinal
& Single
& Longitudinal
& Single
& Longitudinal
\\

\midrule

CNN
& 0.465 $\pm$ 0.012
& 0.563 $\pm$ 0.155
& 0.669 $\pm$ 0.058
& 0.685 $\pm$ 0.236
& 0.461 $\pm$ 0.046
& 0.553 $\pm$ 0.104
\\

ViT
& 0.528 $\pm$ 0.012
& 0.754 $\pm$ 0.075
& 0.621 $\pm$ 0.015
& 0.767 $\pm$ 0.096
& 0.476 $\pm$ 0.071
& 0.720 $\pm$ 0.142
\\

Swin Transformer
& 0.481 $\pm$ 0.039
& 0.693 $\pm$ 0.038
& 0.701 $\pm$ 0.054
& 0.769 $\pm$ 0.049
& 0.484 $\pm$ 0.039
& 0.667 $\pm$ 0.056
\\

\textbf{STRIDE}
& \textbf{0.699 $\pm$ 0.046}
& \textbf{0.816 $\pm$ 0.083}
& \textbf{0.712 $\pm$ 0.058}
& \textbf{0.845 $\pm$ 0.049}
& \textbf{0.748 $\pm$ 0.102}
& \textbf{0.796 $\pm$ 0.072}
\\

\midrule

\textbf{Average}
& 0.543
& \textbf{0.706}
& 0.676
& \textbf{0.767}
& 0.542
& \textbf{0.684}
\\

\bottomrule
\end{tabular}
\end{adjustbox}
\end{table*}


\begin{table*}[!htbp]
\centering
\footnotesize
\setlength{\tabcolsep}{3pt}

\caption{
Complete comparison with longitudinal modeling strategies for three-class
post-treatment classification. Macro-averaged one-vs-rest ROC--AUC,
accuracy, and macro F1 are reported as mean $\pm$ standard deviation across
the ten held-out test folds.
}
\label{tab:app_longitudinal_baselines}

\begin{adjustbox}{max width=\textwidth}
\begin{tabular}{llccc}
\toprule

Model
& Temporal strategy
& Macro ROC--AUC $\uparrow$
& Accuracy $\uparrow$
& Macro F1 $\uparrow$
\\

\midrule

Direct Feature Fusion
& Concatenation / difference
& 0.681 $\pm$ 0.095
& 0.695 $\pm$ 0.082
& 0.628 $\pm$ 0.125
\\

Siamese
& Cross-visit feature comparison
& 0.739 $\pm$ 0.062
& 0.788 $\pm$ 0.105
& 0.754 $\pm$ 0.094
\\

LSTM
& Recurrent temporal aggregation
& 0.656 $\pm$ 0.057
& 0.698 $\pm$ 0.099
& 0.631 $\pm$ 0.077
\\

Temporal Transformer
& Attention-based temporal modeling
& 0.768 $\pm$ 0.049
& 0.813 $\pm$ 0.069
& 0.760 $\pm$ 0.081
\\

\textbf{STRIDE}
& \textbf{Time-conditioned latent transition}
& \textbf{0.816 $\pm$ 0.083}
& \textbf{0.845 $\pm$ 0.049}
& \textbf{0.796 $\pm$ 0.072}
\\

\bottomrule
\end{tabular}
\end{adjustbox}
\end{table*}


\begin{table*}[!htbp]
\centering
\footnotesize
\setlength{\tabcolsep}{3.5pt}

\caption{
Complete cumulative ablation study of STRIDE for three-class post-treatment
classification. Macro-averaged one-vs-rest ROC--AUC, accuracy, and macro F1
are reported as mean $\pm$ standard deviation across the ten held-out test
folds. Dir. Loss denotes latent directional consistency loss and Obs.--Trans.
denotes observed--transition fusion.
}
\label{tab:app_temporal_ablation}

\begin{adjustbox}{max width=\textwidth}
\begin{tabular}{lccccccc}
\toprule

Setting
& Interval
& Transition
& Dir. Loss
& Obs.--Trans.
& \shortstack{ROC--AUC $\uparrow$}
& \shortstack{Accuracy $\uparrow$}
& \shortstack{Macro F1 $\uparrow$}
\\

\midrule

Baseline Fusion
&
&
&
&
& 0.678 $\pm$ 0.095
& 0.695 $\pm$ 0.082
& 0.628 $\pm$ 0.125
\\

+ Time Interval
& \checkmark
&
&
&
& 0.699 $\pm$ 0.033
& 0.726 $\pm$ 0.081
& 0.654 $\pm$ 0.148
\\

+ Latent Transition
& \checkmark
& \checkmark
&
&
& 0.741 $\pm$ 0.092
& 0.769 $\pm$ 0.108
& 0.726 $\pm$ 0.039
\\

+ Directional Consistency
& \checkmark
& \checkmark
& \checkmark
&
& 0.760 $\pm$ 0.077
& 0.783 $\pm$ 0.101
& 0.745 $\pm$ 0.062
\\

\textbf{Full STRIDE}
& \checkmark
& \checkmark
& \checkmark
& \checkmark
& \textbf{0.816 $\pm$ 0.083}
& \textbf{0.845 $\pm$ 0.049}
& \textbf{0.796 $\pm$ 0.072}
\\

\bottomrule
\end{tabular}
\end{adjustbox}
\end{table*}


\begin{table}[!htbp]
\centering
\small
\caption{
Paired comparison of fold-wise macro-averaged one-vs-rest ROC--AUC
on the Burdenko cohort. Performance is reported as mean $\pm$
standard deviation across the ten patient-level test folds.
$\Delta$ is the mean within-fold difference between STRIDE and
the indicated baseline. Adjusted $p$ values account for the four
baseline comparisons.
}
\label{tab:app_paired_auc}

\begin{adjustbox}{max width=\textwidth}
\begin{tabular}{lcccc}
\toprule
Model
& Macro ROC--AUC
& $\Delta$ ROC--AUC
& 95\% CI for $\Delta$
& Holm $p$ \\
\midrule
Direct Feature Fusion
& 0.681 $\pm$ 0.095 & $0.135$ & [0.121, 0.149] & $1.39 \times 10^{-8}$ \\
Siamese
& 0.739 $\pm$ 0.062 & $0.077$ & [0.050, 0.103] & $2.05 \times 10^{-4}$ \\
LSTM
& 0.656 $\pm$ 0.057 & $0.160$ & [0.134, 0.186] & $6.46 \times 10^{-7}$ \\
Temporal Transformer
& 0.768 $\pm$ 0.049 & $0.048$ & [0.023, 0.072] & 0.00144 \\
\midrule
\textbf{STRIDE}
& 0.816 $\pm$ 0.083 & \textemdash & \textemdash & \textemdash \\
\bottomrule
\end{tabular}
\end{adjustbox}
\end{table}


\FloatBarrier


\begin{figure*}[!htbp]
\centering
\includegraphics[width=0.95\linewidth]
{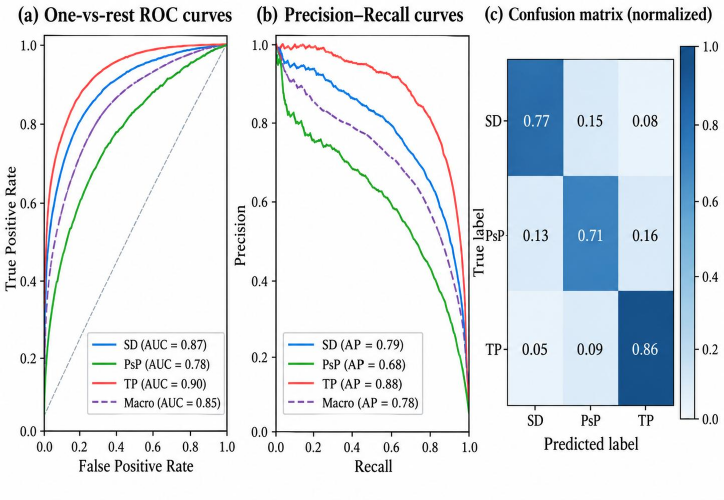}

\caption{
Class-wise performance of STRIDE based on pooled out-of-fold predictions
from the patient-level ten-fold cross-validation.
(a) One-vs-rest receiver operating characteristic (ROC) curves for stable
disease (SD), pseudoprogression (PsP), and true progression (TP). After
aggregating the held-out predictions from all ten test folds, the pooled
out-of-fold macro ROC--AUC was 0.85.
(b) Precision--recall curves computed from the pooled out-of-fold
predictions for the three clinical-state classes.
(c) Normalized confusion matrix computed from the pooled out-of-fold
predictions. The pooled ROC--AUC reported here differs from the
fold-wise macro ROC--AUC of $0.816 \pm 0.083$ reported in the main
manuscript because the former is calculated once after aggregating all
held-out predictions, whereas the latter represents the mean $\pm$
standard deviation of ROC--AUC values calculated independently across
the ten test folds.
}
\label{fig:app_classwise_performance}
\end{figure*}


\FloatBarrier
\clearpage

\bibliographystyle{elsarticle-harv}
\bibliography{references}

\end{document}